\documentclass[conference]{IEEEtran}
\usepackage{cite}

\ifCLASSINFOpdf
\else
\fi
\usepackage{float}		
\usepackage{times}		
\usepackage{graphicx}		
\usepackage{epstopdf}
\usepackage{url}
\usepackage{amsfonts }
\usepackage{amssymb,amsmath}
\usepackage{verbatim}
\usepackage{algorithm}
\usepackage{algorithmic}
\usepackage{amsthm }
\usepackage{amsmath}
\usepackage{tikz}
\usepackage{relsize}

\newcommand{\bmx}{\begin{bmatrix}}
\newcommand{\emx}{\end{bmatrix}}
\newcommand{\bsmx}{\left[\begin{smallmatrix}}
\newcommand{\esmx}{\end{smallmatrix}\right]}

\newcommand{\be}{\begin{equation}}
\newcommand{\ee}{\end{equation}}
\newcommand{\beqy}{\begin{eqnarray}}
\newcommand{\eeqy}{\end{eqnarray}}
\newcommand{\beqynn}{\begin{eqnarray*}}
\newcommand{\eeqynn}{\end{eqnarray*}}

\newcommand{\by}{{\bar{y}}}

\newcommand{\dist}{\mathrm{dist}}

\begin{document}
%
\title{Column Reordering for Box-Constrained Integer Least Squares Problems}

\author{
\IEEEauthorblockN{Stephen Breen}
\IEEEauthorblockA{School of Computer Science\\
McGill University\\
Montreal, Quebec, Canada\\
Email: sbreen1@cs.mcgill.ca}
\and
\IEEEauthorblockN{Xiao-Wen Chang}
\IEEEauthorblockA{School of Computer Science\\
McGill University\\
Montreal, Quebec, Canada\\
Email: chang@cs.mcgill.ca}}


%


\parindent 8pt

\maketitle

\begin{abstract}
The box-constrained integer least squares problem (BILS) arises in MIMO wireless
communications applications. Typically a sphere decoding algorithm (a tree search algorithm)
is used to solve the problem. In order to make the search algorithm
 more efficient, the columns of the channel matrix in the BILS problem have to
be reordered.  
To our knowledge, there are currently two algorithms for column reordering that provide the best known results. 
Both use all available information, but they were derived respectively from geometric and algebraic points of view
and look different. 
In this paper we modify one to make it more computationally efficient and easier to comprehend.
Then we prove the modified one and the other actually give the same column reordering in theory.
Finally we propose a new mathematically equivalent algorithm, which is more computationally efficient
and is still easy to understand.
\end{abstract}


%
\IEEEpeerreviewmaketitle

\section{Introduction}
Given a real vector $y\in\mathbb{R}^m$ and a real matrix $H\in\mathbb{R}^{m \times n}$, 
integer vectors $l, u \in \mathbb{Z}^n$ with $l<u$,
the box-constrained integer least squares (BILS)
problem is defined as:
\be\label{eq:bils0}
\min_{x\in {\cal B }}  \| y- Hx \|_2,  
\ee
where ${\cal B} = {\cal B}_1  \times \cdots \times {\cal B}_n$ with
${\cal B}_i = \{x_i \in\mathbb{Z} : l_i \le x_i \le u_i \}$.
This problem arises in wireless communications applications such as MIMO signal decoding. 
In this paper, we assume that $H$ has full column rank.
The set $\{w=Hx : x\in \mathbb{Z}^n\}$ is referred to as the lattice
generated by $H$.

Let $H$ have the QR factorization
$$
H=[Q_1, Q_2] \bmx R \\ 0 \emx,
$$
where $[\underset{n}{Q_1}, \underset{m-n}{Q_2}]  \in \mathbb{R}^{m\times m}$ is orthogonal
and $R\in \mathbb{R}^{n\times n}$ is upper triangular. 
Then, with $\bar{y}=Q_1^Ty$ 
the BILS problem \eqref{eq:bils0} is reduced to 
\be \label{eq:bils}
 \min_{x \in  {\cal B}}  \| \by- Rx \|_2.
\ee
To solve this reduced problem sphere decoding search algorithms (see, e.g.,  \cite{DamGC03}, \cite{BouGBF03}
and  \cite{ChaH05}) enumerate  the elements in ${\cal B}$ in some order
to find the optimal solution. 

If we reorder the columns of $H$,  i.e., we apply a permutation matrix $P$ to $H$ from the right,
then we will obtain a different R-factor,  resulting in different search speed.
A few algorithms have been proposed to
find $P$ to minimize the complexity of the search
algorithms. In \cite{DamGC03}, the well-known V-BLAST column reordering strategy
originally given in \cite{FosGVW99} was proposed for this purpose.
In \cite{ChaH05}, the SQRD column reordering strategy originally 
presented in \cite{WubBRKK01} for the same purpose as V-BLAST,
was proposed for this purpose.
Both strategies use only the information of the matrix $H$.

In \cite{SuW05}, Su and Wassell considered the geometry of the BILS
problem for the case that $H$ is nonsingular and proposed a new column reordering algorithm (to be called
the SW algorithm from here on for convenience) which uses all information of the BILS problem \eqref{eq:bils0}.
Unfortunately, in our point of view, the geometric interpretation of this algorithm is hard to understand.
Probably due to page limit, the description of the algorithm is very concise, 
making efficient implementation difficult for ordinary users. 

In this paper we will give some new insight of the SW algorithm from an algebraic point of view.
We will make some modifications so that the algorithm becomes more efficient
and easier to understand and furthermore it can handle a general full column rank $H$.

Independently  Chang and Han in \cite{ChaH05} proposed
another column reordering algorithm (which will be referred to as  CH).
Their algorithm also uses all information of \eqref{eq:bils0} and the derivation
is based on an algebraic point of view. It is  easy to see from the equations in
the search process exactly what the CH column reordering is doing and why we
should expect a reduced complexity in the search process. The detailed
description of the CH column reordering is given in \cite{ChaH05} and it is easy
for others to implement the algorithm.
But our numerical tests indicated CH has a  higher complexity than SW, when SW
is implemented efficiently.
Our numerical tests also showed that CH and SW {\em almost} always   
produced the same permutation matrix $P$.

In this paper, we will show that the CH algorithm and the  (modified)  SW algorithm give the same
column reordering in theory. 
This is interesting because both algorithms were derived through different motivations
and we now have both a geometric justification and an algebraic justification 
for why the column reordering strategy should reduce the complexity of the search.
Furthermore, using the knowledge that certain steps in each algorithm are equivalent,
we can combine the best parts from each into a new algorithm. The new algorithm
has a lower flop count than either of the originals.
This is important to the successive interference cancellation decoder, 
which computes a suboptimal solution to \eqref{eq:bils0}.
The new algorithm can be interpreted in the same way as CH,
so it is easy to understand.

In this paper,  $e_i$ denotes the $i^{th}$ column of the identity matrix $I$.
For a set of integer numbers ${\cal S}$ and real number $x$, 
$\lfloor x\rceil_{\cal S}$ denotes the nearest integer in  ${\cal S}$
to $x$ and if there is a tie it denotes the one which has smaller magnitude.
For $z\in{\cal S}$, ${\cal S}\backslash z$ denotes ${\cal S}$ after $z$ is removed.   
We sometimes use MATLAB-like notation for matrices and vectors, e.g.,
$A_{1:m,1:n}$ denotes the matrix formed by the first $m$ rows
and $n$ columns of the matrix $A$ and $A_{:,1:n}$ denote the matrix formed by
the first $n$ columns of $A$.
The $j^{th}$ column of a matrix $A$ is demoted either by $a_j$ or $A_{:,j}$.

\section{Search Process}
\label{sec:Search}
Both CH and SW column reordering algorithms use ideas that arise from the search
process. 
Before the column reorderings are introduced, it is
important to have an understanding of the sphere decoding search process. 

Consider the ILS problem \eqref{eq:bils}. We would like to enumerate the
elements in ${\cal B}$ in an efficient manner in order to find the solution $x$.
One such enumeration strategy is described in \cite{ChaH05}. We will now
describe it briefly.

Suppose that the solution satisfies the following bound, 
\begin{equation}
\left \| \bar{y} - Rx \right \|_2^2 < \beta.
\label{eq:searchIneq}
\end{equation}
There are a few ways to choose a valid initial value for $\beta$, see, e.g.,
\cite{ChaH05}. The inequality (\ref{eq:searchIneq}) defines an ellipsoid in terms of $x$ 
or a hyper-sphere in terms of the lattice point $w=Rx$ with radius $\beta$. 
Define
\begin{equation}
 c_k = (\bar{y}_k - \sum_{j=k+1}^nr_{kj}x_j)/r_{kk}, \; k=n, n-1,\ldots, 1,
\label{eq:searchC}
\end{equation}
where when $k=n$ the sum in the right hand side does not exist.
Then \eqref{eq:searchIneq} can be rewritten as
$$
\sum_{k=1}^n r_{kk}^2(x_k-c_k)^2 < \beta,
$$
which implies the following
set of inequalities:
\begin{align}
\text{level } k: \ \ r_{kk}^2(x_k-c_k)^2 < \beta -
\sum_{i=k+1}^nr_{ii}^2(x_i-c_i)^2, \label{eq:searchLevelK} 
\end{align}
for $k=n,n-1,\ldots, 1$.

We begin the search process
at level $n$. Choose $x_n = \lfloor c_n \rceil_{{\cal B}_n}$, the nearest
integer in ${\cal B}_n$ to $c_n$. If the inequality (\ref{eq:searchLevelK}) with $k=n$
is not satisfied, it will not be satisfied for any integer, this means $\beta$
was chosen to be too small, it must be enlarged. With $x_n$ fixed, we can move
to level $n-1$ and choose $x_{n-1} = \lfloor c_{n-1} \rceil_{{\cal B}_{n-1}}$ with $c_{n-1}$ calculated as in (\ref{eq:searchC}). At this point it
is possible that the inequality (\ref{eq:searchLevelK}) is no longer satisfied.
If this is the case, we must move back to level $n$ and choose $x_n$ to be the
second nearest integer to $c_n$.  We will continue this procedure until we reach
level 1, moving back a level if ever the inequality for the current level is no
longer satisfied.
When we reach level $1$, we will have found an integer point $\hat{x}$. We then
update $\beta = \left \| \bar{y} - R\hat{x} \right \|_2^2$ and try
to find a better integer point which satisfies the box-constraint in the new
ellipsoid. Finally in the search process, when we can no longer find any $x_n$
to satisfy (\ref{eq:searchLevelK}) with $k=n$, the search process is complete and the last integer
point $\hat{x}$ found is the solution. 

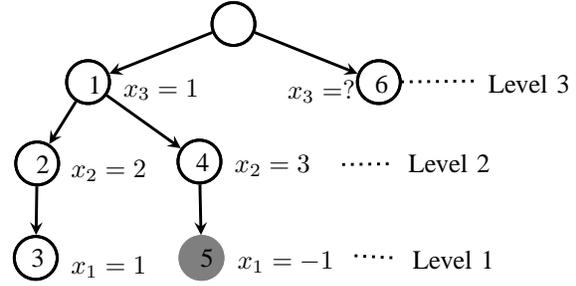
\begin{figure}
\ifx\du\undefined
  \newlength{\du}
\fi
\setlength{\du}{11\unitlength}
\begin{tikzpicture}
\pgftransformxscale{1.000000}
\pgftransformyscale{-1.000000}
\definecolor{dialinecolor}{rgb}{0.000000, 0.000000, 0.000000}
\pgfsetstrokecolor{dialinecolor}
\definecolor{dialinecolor}{rgb}{1.000000, 1.000000, 1.000000}
\pgfsetfillcolor{dialinecolor}
\pgfsetlinewidth{0.100000\du}
\pgfsetdash{}{0pt}
\pgfsetdash{}{0pt}
\pgfsetbuttcap
\pgfsetmiterjoin
\pgfsetlinewidth{0.100000\du}
\pgfsetbuttcap
\pgfsetmiterjoin
\pgfsetdash{}{0pt}
\definecolor{dialinecolor}{rgb}{1.000000, 1.000000, 1.000000}
\pgfsetfillcolor{dialinecolor}
\pgfpathellipse{\pgfpoint{7.737500\du}{1.737500\du}}{\pgfpoint{0.737500\du}{0\du
}}{\pgfpoint{0\du}{0.737500\du}}
\pgfusepath{fill}
\definecolor{dialinecolor}{rgb}{0.000000, 0.000000, 0.000000}
\pgfsetstrokecolor{dialinecolor}
\pgfpathellipse{\pgfpoint{7.737500\du}{1.737500\du}}{\pgfpoint{0.737500\du}{0\du
}}{\pgfpoint{0\du}{0.737500\du}}
\pgfusepath{stroke}
\pgfsetbuttcap
\pgfsetmiterjoin
\pgfsetdash{}{0pt}
\definecolor{dialinecolor}{rgb}{0.000000, 0.000000, 0.000000}
\pgfsetstrokecolor{dialinecolor}
\pgfpathellipse{\pgfpoint{7.737500\du}{1.737500\du}}{\pgfpoint{0.737500\du}{0\du
}}{\pgfpoint{0\du}{0.737500\du}}
\pgfusepath{stroke}
\pgfsetlinewidth{0.100000\du}
\pgfsetdash{}{0pt}
\pgfsetdash{}{0pt}
\pgfsetbuttcap
\pgfsetmiterjoin
\pgfsetlinewidth{0.100000\du}
\pgfsetbuttcap
\pgfsetmiterjoin
\pgfsetdash{}{0pt}
\definecolor{dialinecolor}{rgb}{1.000000, 1.000000, 1.000000}
\pgfsetfillcolor{dialinecolor}
\pgfpathellipse{\pgfpoint{2.737500\du}{3.737500\du}}{\pgfpoint{0.737500\du}{0\du
}}{\pgfpoint{0\du}{0.737500\du}}
\pgfusepath{fill}
\definecolor{dialinecolor}{rgb}{0.000000, 0.000000, 0.000000}
\pgfsetstrokecolor{dialinecolor}
\pgfpathellipse{\pgfpoint{2.737500\du}{3.737500\du}}{\pgfpoint{0.737500\du}{0\du
}}{\pgfpoint{0\du}{0.737500\du}}
\pgfusepath{stroke}
\pgfsetbuttcap
\pgfsetmiterjoin
\pgfsetdash{}{0pt}
\definecolor{dialinecolor}{rgb}{0.000000, 0.000000, 0.000000}
\pgfsetstrokecolor{dialinecolor}
\pgfpathellipse{\pgfpoint{2.737500\du}{3.737500\du}}{\pgfpoint{0.737500\du}{0\du
}}{\pgfpoint{0\du}{0.737500\du}}
\pgfusepath{stroke}
\pgfsetlinewidth{0.100000\du}
\pgfsetdash{}{0pt}
\pgfsetdash{}{0pt}
\pgfsetbuttcap
\pgfsetmiterjoin
\pgfsetlinewidth{0.100000\du}
\pgfsetbuttcap
\pgfsetmiterjoin
\pgfsetdash{}{0pt}
\definecolor{dialinecolor}{rgb}{1.000000, 1.000000, 1.000000}
\pgfsetfillcolor{dialinecolor}
\pgfpathellipse{\pgfpoint{12.737500\du}{3.737500\du}}{\pgfpoint{0.737500\du}{
0\du}}{\pgfpoint{0\du}{0.737500\du}}
\pgfusepath{fill}
\definecolor{dialinecolor}{rgb}{0.000000, 0.000000, 0.000000}
\pgfsetstrokecolor{dialinecolor}
\pgfpathellipse{\pgfpoint{12.737500\du}{3.737500\du}}{\pgfpoint{0.737500\du}{
0\du}}{\pgfpoint{0\du}{0.737500\du}}
\pgfusepath{stroke}
\pgfsetbuttcap
\pgfsetmiterjoin
\pgfsetdash{}{0pt}
\definecolor{dialinecolor}{rgb}{0.000000, 0.000000, 0.000000}
\pgfsetstrokecolor{dialinecolor}
\pgfpathellipse{\pgfpoint{12.737500\du}{3.737500\du}}{\pgfpoint{0.737500\du}{
0\du}}{\pgfpoint{0\du}{0.737500\du}}
\pgfusepath{stroke}
\pgfsetlinewidth{0.100000\du}
\pgfsetdash{}{0pt}
\pgfsetdash{}{0pt}
\pgfsetbuttcap
\pgfsetmiterjoin
\pgfsetlinewidth{0.100000\du}
\pgfsetbuttcap
\pgfsetmiterjoin
\pgfsetdash{}{0pt}
\definecolor{dialinecolor}{rgb}{1.000000, 1.000000, 1.000000}
\pgfsetfillcolor{dialinecolor}
\pgfpathellipse{\pgfpoint{0.987500\du}{6.537500\du}}{\pgfpoint{0.737500\du}{0\du
}}{\pgfpoint{0\du}{0.737500\du}}
\pgfusepath{fill}
\definecolor{dialinecolor}{rgb}{0.000000, 0.000000, 0.000000}
\pgfsetstrokecolor{dialinecolor}
\pgfpathellipse{\pgfpoint{0.987500\du}{6.537500\du}}{\pgfpoint{0.737500\du}{0\du
}}{\pgfpoint{0\du}{0.737500\du}}
\pgfusepath{stroke}
\pgfsetbuttcap
\pgfsetmiterjoin
\pgfsetdash{}{0pt}
\definecolor{dialinecolor}{rgb}{0.000000, 0.000000, 0.000000}
\pgfsetstrokecolor{dialinecolor}
\pgfpathellipse{\pgfpoint{0.987500\du}{6.537500\du}}{\pgfpoint{0.737500\du}{0\du
}}{\pgfpoint{0\du}{0.737500\du}}
\pgfusepath{stroke}
\pgfsetlinewidth{0.100000\du}
\pgfsetdash{}{0pt}
\pgfsetdash{}{0pt}
\pgfsetbuttcap
\pgfsetmiterjoin
\pgfsetlinewidth{0.100000\du}
\pgfsetbuttcap
\pgfsetmiterjoin
\pgfsetdash{}{0pt}
\definecolor{dialinecolor}{rgb}{0.498039, 0.498039, 0.498039}
\pgfsetfillcolor{dialinecolor}
\pgfpathellipse{\pgfpoint{6.637500\du}{9.787500\du}}{\pgfpoint{0.737500\du}{0\du
}}{\pgfpoint{0\du}{0.737500\du}}
\pgfusepath{fill}
\definecolor{dialinecolor}{rgb}{0.498039, 0.498039, 0.498039}
\pgfsetstrokecolor{dialinecolor}
\pgfpathellipse{\pgfpoint{6.637500\du}{9.787500\du}}{\pgfpoint{0.737500\du}{0\du
}}{\pgfpoint{0\du}{0.737500\du}}
\pgfusepath{stroke}
\pgfsetbuttcap
\pgfsetmiterjoin
\pgfsetdash{}{0pt}
\definecolor{dialinecolor}{rgb}{0.498039, 0.498039, 0.498039}
\pgfsetstrokecolor{dialinecolor}
\pgfpathellipse{\pgfpoint{6.637500\du}{9.787500\du}}{\pgfpoint{0.737500\du}{0\du
}}{\pgfpoint{0\du}{0.737500\du}}
\pgfusepath{stroke}
\pgfsetlinewidth{0.100000\du}
\pgfsetdash{}{0pt}
\pgfsetdash{}{0pt}
\pgfsetbuttcap
\pgfsetmiterjoin
\pgfsetlinewidth{0.100000\du}
\pgfsetbuttcap
\pgfsetmiterjoin
\pgfsetdash{}{0pt}
\definecolor{dialinecolor}{rgb}{1.000000, 1.000000, 1.000000}
\pgfsetfillcolor{dialinecolor}
\pgfpathellipse{\pgfpoint{0.937500\du}{9.787500\du}}{\pgfpoint{0.737500\du}{0\du
}}{\pgfpoint{0\du}{0.737500\du}}
\pgfusepath{fill}
\definecolor{dialinecolor}{rgb}{0.000000, 0.000000, 0.000000}
\pgfsetstrokecolor{dialinecolor}
\pgfpathellipse{\pgfpoint{0.937500\du}{9.787500\du}}{\pgfpoint{0.737500\du}{0\du
}}{\pgfpoint{0\du}{0.737500\du}}
\pgfusepath{stroke}
\pgfsetbuttcap
\pgfsetmiterjoin
\pgfsetdash{}{0pt}
\definecolor{dialinecolor}{rgb}{0.000000, 0.000000, 0.000000}
\pgfsetstrokecolor{dialinecolor}
\pgfpathellipse{\pgfpoint{0.937500\du}{9.787500\du}}{\pgfpoint{0.737500\du}{0\du
}}{\pgfpoint{0\du}{0.737500\du}}
\pgfusepath{stroke}
\pgfsetlinewidth{0.100000\du}
\pgfsetdash{}{0pt}
\pgfsetdash{}{0pt}
\pgfsetbuttcap
{
\definecolor{dialinecolor}{rgb}{0.000000, 0.000000, 0.000000}
\pgfsetfillcolor{dialinecolor}
\pgfsetarrowsend{stealth}
\definecolor{dialinecolor}{rgb}{0.000000, 0.000000, 0.000000}
\pgfsetstrokecolor{dialinecolor}
\draw (7.006909\du,2.029736\du)--(3.468091\du,3.445264\du);
}
\pgfsetlinewidth{0.100000\du}
\pgfsetdash{}{0pt}
\pgfsetdash{}{0pt}
\pgfsetbuttcap
{
\definecolor{dialinecolor}{rgb}{0.000000, 0.000000, 0.000000}
\pgfsetfillcolor{dialinecolor}
\pgfsetarrowsend{stealth}
\definecolor{dialinecolor}{rgb}{0.000000, 0.000000, 0.000000}
\pgfsetstrokecolor{dialinecolor}
\draw (8.468091\du,2.029736\du)--(12.006909\du,3.445264\du);
}
\pgfsetlinewidth{0.100000\du}
\pgfsetdash{}{0pt}
\pgfsetdash{}{0pt}
\pgfsetbuttcap
{
\definecolor{dialinecolor}{rgb}{0.000000, 0.000000, 0.000000}
\pgfsetfillcolor{dialinecolor}
\pgfsetarrowsend{stealth}
\definecolor{dialinecolor}{rgb}{0.000000, 0.000000, 0.000000}
\pgfsetstrokecolor{dialinecolor}
\draw (2.320081\du,4.405371\du)--(1.404919\du,5.869629\du);
}
\pgfsetlinewidth{0.100000\du}
\pgfsetdash{}{0pt}
\pgfsetdash{}{0pt}
\pgfsetbuttcap
{
\definecolor{dialinecolor}{rgb}{0.000000, 0.000000, 0.000000}
\pgfsetfillcolor{dialinecolor}
\pgfsetarrowsend{stealth}
\definecolor{dialinecolor}{rgb}{0.000000, 0.000000, 0.000000}
\pgfsetstrokecolor{dialinecolor}
\draw (0.975385\du,7.325006\du)--(0.949615\du,8.999994\du);
}
\pgfsetlinewidth{0.100000\du}
\pgfsetdash{}{0pt}
\pgfsetdash{}{0pt}
\pgfsetbuttcap
{
\definecolor{dialinecolor}{rgb}{0.000000, 0.000000, 0.000000}
\pgfsetfillcolor{dialinecolor}
\pgfsetarrowsend{stealth}
\definecolor{dialinecolor}{rgb}{0.000000, 0.000000, 0.000000}
\pgfsetstrokecolor{dialinecolor}
\draw (6.584128\du,7.259974\du)--(6.620872\du,9.000026\du);
}
\definecolor{dialinecolor}{rgb}{0.000000, 0.000000, 0.000000}
\pgfsetstrokecolor{dialinecolor}
\node[anchor=west] at (3.600000\du,4.000000\du){$x_3 = 1$};
\definecolor{dialinecolor}{rgb}{0.000000, 0.000000, 0.000000}
\pgfsetstrokecolor{dialinecolor}
\node[anchor=west] at (9.250000\du,4.050000\du){$x_3 = ?$};
\definecolor{dialinecolor}{rgb}{0.000000, 0.000000, 0.000000}
\pgfsetstrokecolor{dialinecolor}
\node[anchor=west] at (1.800000\du,6.800000\du){$x_{2}=2$};
\definecolor{dialinecolor}{rgb}{0.000000, 0.000000, 0.000000}
\pgfsetstrokecolor{dialinecolor}
\node[anchor=west] at (1.800000\du,10.100000\du){$x_{1}=1$};
\definecolor{dialinecolor}{rgb}{0.000000, 0.000000, 0.000000}
\pgfsetstrokecolor{dialinecolor}
\node[anchor=west] at (7.530000\du,9.935000\du){$x_{1}=-1$};
\pgfsetlinewidth{0.100000\du}
\pgfsetdash{{\pgflinewidth}{0.200000\du}}{0cm}
\pgfsetdash{{\pgflinewidth}{0.200000\du}}{0cm}
\pgfsetbuttcap
{
\definecolor{dialinecolor}{rgb}{0.000000, 0.000000, 0.000000}
\pgfsetfillcolor{dialinecolor}
\definecolor{dialinecolor}{rgb}{0.000000, 0.000000, 0.000000}
\pgfsetstrokecolor{dialinecolor}
\draw (13.499316\du,3.729004\du)--(16.100000\du,3.700000\du);
}
\pgfsetlinewidth{0.100000\du}
\pgfsetdash{{\pgflinewidth}{0.200000\du}}{0cm}
\pgfsetdash{{\pgflinewidth}{0.200000\du}}{0cm}
\pgfsetbuttcap
{
\definecolor{dialinecolor}{rgb}{0.000000, 0.000000, 0.000000}
\pgfsetfillcolor{dialinecolor}
\definecolor{dialinecolor}{rgb}{0.000000, 0.000000, 0.000000}
\pgfsetstrokecolor{dialinecolor}
\draw (11.474979\du,6.575259\du)--(13.100000\du,6.550000\du);
}
\pgfsetlinewidth{0.100000\du}
\pgfsetdash{{\pgflinewidth}{0.200000\du}}{0cm}
\pgfsetdash{{\pgflinewidth}{0.200000\du}}{0cm}
\pgfsetbuttcap
{
\definecolor{dialinecolor}{rgb}{0.000000, 0.000000, 0.000000}
\pgfsetfillcolor{dialinecolor}
\definecolor{dialinecolor}{rgb}{0.000000, 0.000000, 0.000000}
\pgfsetstrokecolor{dialinecolor}
\draw (11.902902\du,9.713884\du)--(13.300000\du,9.787500\du);
}
\pgfsetlinewidth{0.100000\du}
\pgfsetdash{}{0pt}
\pgfsetdash{}{0pt}
\pgfsetbuttcap
\pgfsetmiterjoin
\pgfsetlinewidth{0.100000\du}
\pgfsetbuttcap
\pgfsetmiterjoin
\pgfsetdash{}{0pt}
\definecolor{dialinecolor}{rgb}{1.000000, 1.000000, 1.000000}
\pgfsetfillcolor{dialinecolor}
\pgfpathellipse{\pgfpoint{6.567500\du}{6.472500\du}}{\pgfpoint{0.737500\du}{0\du
}}{\pgfpoint{0\du}{0.737500\du}}
\pgfusepath{fill}
\definecolor{dialinecolor}{rgb}{0.000000, 0.000000, 0.000000}
\pgfsetstrokecolor{dialinecolor}
\pgfpathellipse{\pgfpoint{6.567500\du}{6.472500\du}}{\pgfpoint{0.737500\du}{0\du
}}{\pgfpoint{0\du}{0.737500\du}}
\pgfusepath{stroke}
\pgfsetbuttcap
\pgfsetmiterjoin
\pgfsetdash{}{0pt}
\definecolor{dialinecolor}{rgb}{0.000000, 0.000000, 0.000000}
\pgfsetstrokecolor{dialinecolor}
\pgfpathellipse{\pgfpoint{6.567500\du}{6.472500\du}}{\pgfpoint{0.737500\du}{0\du
}}{\pgfpoint{0\du}{0.737500\du}}
\pgfusepath{stroke}
\pgfsetlinewidth{0.100000\du}
\pgfsetdash{}{0pt}
\pgfsetdash{}{0pt}
\pgfsetbuttcap
{
\definecolor{dialinecolor}{rgb}{0.000000, 0.000000, 0.000000}
\pgfsetfillcolor{dialinecolor}
\pgfsetarrowsend{stealth}
\definecolor{dialinecolor}{rgb}{0.000000, 0.000000, 0.000000}
\pgfsetstrokecolor{dialinecolor}
\draw (3.378015\du,4.194891\du)--(5.926985\du,6.015109\du);
}
\definecolor{dialinecolor}{rgb}{0.000000, 0.000000, 0.000000}
\pgfsetstrokecolor{dialinecolor}
\node[anchor=west] at (7.430000\du,6.617500\du){$x_2=3$};
\definecolor{dialinecolor}{rgb}{0.000000, 0.000000, 0.000000}
\pgfsetstrokecolor{dialinecolor}
\node[anchor=west] at (2.400000\du,3.837500\du){1};
\definecolor{dialinecolor}{rgb}{0.000000, 0.000000, 0.000000}
\pgfsetstrokecolor{dialinecolor}
\node[anchor=west] at (0.600000\du,6.550000\du){2};
\definecolor{dialinecolor}{rgb}{0.000000, 0.000000, 0.000000}
\pgfsetstrokecolor{dialinecolor}
\node[anchor=west] at (0.437500\du,9.787500\du){3};
\definecolor{dialinecolor}{rgb}{0.000000, 0.000000, 0.000000}
\pgfsetstrokecolor{dialinecolor}
\node[anchor=west] at (6.100000\du,6.487500\du){4};
\definecolor{dialinecolor}{rgb}{0.000000, 0.000000, 0.000000}
\pgfsetstrokecolor{dialinecolor}
\node[anchor=west] at (12.250000\du,3.787500\du){6};
\definecolor{dialinecolor}{rgb}{1.000000, 1.000000, 1.000000}
\pgfsetstrokecolor{dialinecolor}
\node[anchor=west] at (6.250000\du,9.787500\du){5};
\definecolor{dialinecolor}{rgb}{0.000000, 0.000000, 0.000000}
\pgfsetstrokecolor{dialinecolor}
\node[anchor=west] at (16.150000\du,3.800000\du){Level 3};
\definecolor{dialinecolor}{rgb}{0.000000, 0.000000, 0.000000}
\pgfsetstrokecolor{dialinecolor}
\node[anchor=west] at (13.400000\du,6.550000\du){Level 2};
\definecolor{dialinecolor}{rgb}{0.000000, 0.000000, 0.000000}
\pgfsetstrokecolor{dialinecolor}
\node[anchor=west] at (13.550000\du,9.850000\du){Level 1};
\definecolor{dialinecolor}{rgb}{0.000000, 0.000000, 0.000000}
\pgfsetstrokecolor{dialinecolor}
\node[anchor=west] at (17.100000\du,3.650000\du){};
\definecolor{dialinecolor}{rgb}{0.000000, 0.000000, 0.000000}
\pgfsetstrokecolor{dialinecolor}
\node[anchor=west] at (18.600000\du,6.537500\du){};
\definecolor{dialinecolor}{rgb}{0.000000, 0.000000, 0.000000}
\pgfsetstrokecolor{dialinecolor}
\node[anchor=west] at (7.600000\du,8.050000\du){};
\definecolor{dialinecolor}{rgb}{0.000000, 0.000000, 0.000000}
\pgfsetstrokecolor{dialinecolor}
\node[anchor=west] at (9.100000\du,7.850000\du){};
\end{tikzpicture}

\caption{An example of the search process with solution $x = [-1,3,1]^T$.}
\label{fig:treeSearch}
\end{figure}

The above search process is actually a depth-first tree search, see Fig.\ \ref{fig:treeSearch},
where the number  in a node  denote the step number at which the node is encountered.

\section{Column Reordering}
In this section we  introduce the two orginal column reordering algorithms, CH and
SW and explain their motivations.
We give some new insight on SW and propose a modified version.
We also give a complexity analysis for both algorithms.

\subsection{Chang and Han's Algorithm}
The CH algorithm first computes the QR factorization $H$,
then  tries to reorder the columns of $R$.
The motivation for this algorithm comes from observing equation (\ref{eq:searchLevelK}).
If the inequality is false we know that the current choice for the value of
$x_k$ given $x_{k+1:n}$ are fixed is incorrect and we prune the search tree. We
would like to choose the column permutations so that it is likely that the
inequality will be false at higher levels in the search tree. The CH column reordering strategy
does this by trying to maximize the left hand side of  (\ref{eq:searchLevelK}) with large values of $\left | r_{kk}
\right |$ and minimize the right hand side by making $\left | r_{kk}(x_k-c_k) \right |$
large for values of $k = n,n-1, \dots, 1$.

Here we describe  step 1 of the CH algorithm, which determines the last column of the final $R$ 
(or equivalently the last column of the final $H$).
Subsequent steps are the same but are applied to a subproblem that is one dimension smaller. 
In step 1, for $i = 1,\dots,n$ we interchange
columns $i$ and $n$ of  $R$ (thus entries of $i$ and $n$ in $x$ are also swapped), then return $R$ to upper-triangular
by a series of Givens rotations applied to $R$ from the left, which  are also applied to $\bar{y}$.
To avoid confusion, we denote the new $R$ by $\hat{R}$ and the new $\bar{y}$ by $\hat{y}$.
We then compute  $c_n=\hat{y}_n/\hat{r}_{n,n}$ and 
\be
x_i^c=\arg\min_{x_i\in {\cal B}_i}\left |\hat{r}_{nn}(x_i- c_n) \right | = \left \lfloor c_n \right \rceil_{{\cal B}_i},
\label{eq:xic}
\ee
where the superscript $c$ denotes the CH algorithm. 
Let $\bar{x}_i^c$ be the second closest integer in ${{\cal B}_i}$ to $c_n$,
i.e.,  
$\bar{x}_i^c= \left \lfloor c_n \right \rceil_{{\cal B}_i\backslash x_i^c}.$
Define
\be
\dist_i^c = |\hat{r}_{nn}( \bar{x}_i^c -c_n) |, 
\label{eq:dic}
\ee
which represents the partial residual given when $x_i$ is taken to be $\bar{x}_i^c$.
Let $j = {\arg\max}_i \dist_i^c$.
Then  column $j$ of the original $R$ is chosen to be the $n^{th}$ column of the final $R$.
With the corresponding updated upper triangular $R$ and $\bar{y}$
(here for convenience we have removed hats),
the algorithm then updates  $\bar{y}_{1:n-1}$ again
by setting $\bar{y}_{1:n-1}: = \bar{y}_{1:n-1} - r_{1:n-1,n}x_j$ where $x_j=x_j^c$. 
Choosing  $x_j$ to be  $x_j^c$ here is exactly the same as what the search process does.
We then continue to work on the subproblem 
\be
\min_{\tilde{x}\in \mathbb{Z}^{n-1}} \left \| \bar{y}_{1:n-1}-R_{1:n-1,1:n-1}\tilde{x} \right \|_2,
\label{eq:subc}
\ee
where $\tilde{x}=[x_1,\ldots, x_{j-1}, x_n, x_{j+1}, \ldots x_{n-1}]^T$ satisfies the corresponding box constraint.
The pseudocode of the CH algorithm is given in Algorithm \ref{alg:CH}.

To determine the last column, CH finds the permutation to 
maximize $\left |r_{nn}(\bar{x}_i^c-c_n) \right |$. Using $\bar{x}_i^c$ instead of $x_i^c$
ensures that $\left | \bar{x}_i^c-c_n \right |$ is never less than $0.5$ but
also not very large. This means that usually if $\left | r_{nn}(\bar{x}_i^c-c_n)
\right |$ is large, $\left | r_{nn} \right |$ is large as well and the
requirement to have large $|r_{nn}|$ 
is met.
Using $x_i^c$ would not be a good choice because $\left | x_i^c - c_n \right |$ might be 
very small or even $0$, then column $i$ would not be chosen to be column $n$
even if the corresponding $|r_{nn}|$ is large and on the contrary a column with small $|r_{nn}|$
but large $|x_i^c-c_n|$ may be chosen.

Now we will consider the complexity of CH. The 
significant cost comes from line \ref{l:chg} in Algorithm \ref{alg:CH},
which requires $6(k-i)^2$ flops.
If we sum this cost over all loop iterations and add the cost of the QR factorization by Householder transformations, 
we  get a total complexity of $0.5n^4+2mn^2$ flops.

\begin{algorithm}
\caption{CH Algorithm - Returns $p$, the column permutation vector}
\label{alg:CH}
\begin{algorithmic}[1]
\STATE $p := 1:n$
\STATE $p' := 1:n$
\STATE Compute the QR factorization of $H$: $\bsmx Q_1^T \\ Q_2^T \esmx H= \bsmx R\\ 0 \esmx$
             and compute  $\bar{y} : = Q_1^Ty$
  \FOR{$k=n$ to $2$}
  	\STATE $maxDist := -1$
    \FOR{$i=1$ to $k$}
    	\STATE $\hat{y} := \bar{y}_{1:k}$
    	\STATE $\hat{R} := R_{1:k,1:k}$
        \STATE  \label{l:chg} swap columns $i$ and $k$ of $\hat{R}$, return it  to upper
triangular with Givens rotations, also apply the Givens rotations to $\hat{y}$ 
        \STATE $x_i^c := \left \lfloor\hat{y}_k/\hat{r}_{k,k}\right
\rceil_{{\cal B}_i}$
        \STATE $\bar{x}_i^c := \left \lfloor\hat{y}_k/\hat{r}_{k,k}\right
\rceil_{{\cal B}_i\backslash x_i^c}$
        \STATE $dist_i^c := \left | \hat{r}_{k,k}\bar{x}_i^c - \hat{y}_k
\right | $
        \IF{$dist_i^c > maxDist$}
        	\STATE $maxDist := dist_i^c$
        	\STATE $j := i$
        	\STATE $R' := \hat{R}$
        	\STATE $y' := \hat{y}$
        \ENDIF
    \ENDFOR
    \STATE $p_k := p'_j$
    \STATE Interchange the intervals ${{\cal B}_k}$ and ${{\cal B}_{j}}$
    \STATE Intechange entries $k$ and $j$ in $p'$
    \STATE $R_{1:k,1:k} := R'$
    \STATE $\bar{y}_{1:k} := y' - R'_{1:k,k}x_j^c$
  \ENDFOR
  \STATE $p_1 := p'_1$
\end{algorithmic}
\end{algorithm}

\subsection{Su and Wassell's Algorithm}
The motivation for the SW algorithm comes from examining the geometry of the search process.

\ifx\du\undefined
  \newlength{\du}
\fi
\begin{figure}
\setlength{\du}{8.5\unitlength}
\begin{tikzpicture}
\pgftransformxscale{1.000000}
\pgftransformyscale{-1.000000}
\definecolor{dialinecolor}{rgb}{0.000000, 0.000000, 0.000000}
\pgfsetstrokecolor{dialinecolor}
\definecolor{dialinecolor}{rgb}{1.000000, 1.000000, 1.000000}
\pgfsetfillcolor{dialinecolor}
\pgfsetlinewidth{0.100000\du}
\pgfsetdash{}{0pt}
\pgfsetdash{}{0pt}
\pgfsetmiterjoin
\pgfsetbuttcap
\definecolor{dialinecolor}{rgb}{1.000000, 1.000000, 1.000000}
\pgfsetfillcolor{dialinecolor}
\fill
(6.000000\du,5.000000\du)--(10.972140\du,3.581250\du)--(8.972140\du,
9.768750\du)--(4.000000\du,11.000000\du)--cycle;
\definecolor{dialinecolor}{rgb}{0.000000, 0.000000, 0.000000}
\pgfsetstrokecolor{dialinecolor}
\draw
(6.000000\du,5.000000\du)--(10.972140\du,3.581250\du)--(8.972140\du,
9.768750\du)--(4.000000\du,11.000000\du)--cycle;
\pgfsetlinewidth{0.100000\du}
\pgfsetdash{{\pgflinewidth}{0.200000\du}}{0cm}
\pgfsetdash{{\pgflinewidth}{0.200000\du}}{0cm}
\pgfsetbuttcap
{
\definecolor{dialinecolor}{rgb}{0.000000, 0.000000, 0.000000}
\pgfsetfillcolor{dialinecolor}
\definecolor{dialinecolor}{rgb}{0.000000, 0.000000, 0.000000}
\pgfsetstrokecolor{dialinecolor}
\draw (7.000000\du,2.000000\du)--(2.150000\du,16.550000\du);
}
\pgfsetlinewidth{0.100000\du}
\pgfsetdash{{\pgflinewidth}{0.200000\du}}{0cm}
\pgfsetdash{{\pgflinewidth}{0.200000\du}}{0cm}
\pgfsetbuttcap
{
\definecolor{dialinecolor}{rgb}{0.000000, 0.000000, 0.000000}
\pgfsetfillcolor{dialinecolor}
\definecolor{dialinecolor}{rgb}{0.000000, 0.000000, 0.000000}
\pgfsetstrokecolor{dialinecolor}
\draw (11.847140\du,0.865625\du)--(6.547140\du,17.212500\du);
}
\pgfsetlinewidth{0.050000\du}
\pgfsetdash{}{0pt}
\pgfsetdash{}{0pt}
\pgfsetbuttcap
\pgfsetmiterjoin
\pgfsetlinewidth{0.050000\du}
\pgfsetbuttcap
\pgfsetmiterjoin
\pgfsetdash{}{0pt}
\definecolor{dialinecolor}{rgb}{0.000000, 0.000000, 0.000000}
\pgfsetfillcolor{dialinecolor}
\pgfpathellipse{\pgfpoint{6.411748\du}{10.802248\du}}{\pgfpoint{0.111748\du}{
0\du}}{\pgfpoint{0\du}{0.111748\du}}
\pgfusepath{fill}
\definecolor{dialinecolor}{rgb}{0.000000, 0.000000, 0.000000}
\pgfsetstrokecolor{dialinecolor}
\pgfpathellipse{\pgfpoint{6.411748\du}{10.802248\du}}{\pgfpoint{0.111748\du}{
0\du}}{\pgfpoint{0\du}{0.111748\du}}
\pgfusepath{stroke}
\pgfsetbuttcap
\pgfsetmiterjoin
\pgfsetdash{}{0pt}
\definecolor{dialinecolor}{rgb}{0.000000, 0.000000, 0.000000}
\pgfsetstrokecolor{dialinecolor}
\pgfpathellipse{\pgfpoint{6.411748\du}{10.802248\du}}{\pgfpoint{0.111748\du}{
0\du}}{\pgfpoint{0\du}{0.111748\du}}
\pgfusepath{stroke}
\pgfsetlinewidth{0.100000\du}
\pgfsetdash{}{0pt}
\pgfsetdash{}{0pt}
\pgfsetbuttcap
\pgfsetmiterjoin
\pgfsetlinewidth{0.100000\du}
\pgfsetbuttcap
\pgfsetmiterjoin
\pgfsetdash{}{0pt}
\definecolor{dialinecolor}{rgb}{1.000000, 1.000000, 1.000000}
\pgfsetfillcolor{dialinecolor}
\pgfpathellipse{\pgfpoint{3.982410\du}{10.972410\du}}{\pgfpoint{0.127410\du}{
0\du}}{\pgfpoint{0\du}{0.127410\du}}
\pgfusepath{fill}
\definecolor{dialinecolor}{rgb}{0.000000, 0.000000, 0.000000}
\pgfsetstrokecolor{dialinecolor}
\pgfpathellipse{\pgfpoint{3.982410\du}{10.972410\du}}{\pgfpoint{0.127410\du}{
0\du}}{\pgfpoint{0\du}{0.127410\du}}
\pgfusepath{stroke}
\pgfsetbuttcap
\pgfsetmiterjoin
\pgfsetdash{}{0pt}
\definecolor{dialinecolor}{rgb}{0.000000, 0.000000, 0.000000}
\pgfsetstrokecolor{dialinecolor}
\pgfpathellipse{\pgfpoint{3.982410\du}{10.972410\du}}{\pgfpoint{0.127410\du}{
0\du}}{\pgfpoint{0\du}{0.127410\du}}
\pgfusepath{stroke}
\definecolor{dialinecolor}{rgb}{0.000000, 0.000000, 0.000000}
\pgfsetstrokecolor{dialinecolor}
\node[anchor=west] at (3.537500\du,2.450000\du){$\mathsmaller{F_2(1)}$};
\definecolor{dialinecolor}{rgb}{0.000000, 0.000000, 0.000000}
\pgfsetstrokecolor{dialinecolor}
\node[anchor=west] at (8.340000\du,1.745000\du){$\mathsmaller{F_2(-1)}$};
\pgfsetlinewidth{0.050000\du}
\pgfsetdash{}{0pt}
\pgfsetdash{}{0pt}
\pgfsetbuttcap
{
\definecolor{dialinecolor}{rgb}{0.000000, 0.000000, 0.000000}
\pgfsetfillcolor{dialinecolor}
\definecolor{dialinecolor}{rgb}{0.000000, 0.000000, 0.000000}
\pgfsetstrokecolor{dialinecolor}
\draw (9.687500\du,11.775000\du)--(3.287500\du,9.825000\du);
}
\definecolor{dialinecolor}{rgb}{0.498039, 0.498039, 0.498039}
\pgfsetfillcolor{dialinecolor}
\pgfpathellipse{\pgfpoint{4.322140\du}{10.121875\du}}{\pgfpoint{0.150000\du}{
0\du}}{\pgfpoint{0\du}{0.150000\du}}
\pgfusepath{fill}
\pgfsetlinewidth{0.100000\du}
\pgfsetdash{}{0pt}
\pgfsetdash{}{0pt}
\definecolor{dialinecolor}{rgb}{0.498039, 0.498039, 0.498039}
\pgfsetstrokecolor{dialinecolor}
\pgfpathellipse{\pgfpoint{4.322140\du}{10.121875\du}}{\pgfpoint{0.150000\du}{
0\du}}{\pgfpoint{0\du}{0.150000\du}}
\pgfusepath{stroke}
\definecolor{dialinecolor}{rgb}{0.498039, 0.498039, 0.498039}
\pgfsetfillcolor{dialinecolor}
\pgfpathellipse{\pgfpoint{8.434640\du}{11.396875\du}}{\pgfpoint{0.150000\du}{
0\du}}{\pgfpoint{0\du}{0.150000\du}}
\pgfusepath{fill}
\pgfsetlinewidth{0.100000\du}
\pgfsetdash{}{0pt}
\pgfsetdash{}{0pt}
\definecolor{dialinecolor}{rgb}{0.498039, 0.498039, 0.498039}
\pgfsetstrokecolor{dialinecolor}
\pgfpathellipse{\pgfpoint{8.434640\du}{11.396875\du}}{\pgfpoint{0.150000\du}{
0\du}}{\pgfpoint{0\du}{0.150000\du}}
\pgfusepath{stroke}
\definecolor{dialinecolor}{rgb}{0.000000, 0.000000, 0.000000}
\pgfsetstrokecolor{dialinecolor}
\node[anchor=west] at
(7.297140\du,12.046875\du){$\mathsmaller{\mbox{\smaller{proj}}_{F_2(-1)}(y)}$};
\definecolor{dialinecolor}{rgb}{0.000000, 0.000000, 0.000000}
\pgfsetstrokecolor{dialinecolor}
\node[anchor=west] at
(3.509640\du,9.684375\du){$\mathsmaller{\mbox{\smaller{proj}}_{F_2(1)}(y)}$};
\definecolor{dialinecolor}{rgb}{0.000000, 0.000000, 0.000000}
\pgfsetstrokecolor{dialinecolor}
\node[anchor=west] at (5.806250\du,11.443800\du){$\mathsmaller{y}$};
\pgfsetlinewidth{0.100000\du}
\pgfsetdash{}{0pt}
\pgfsetdash{}{0pt}
\pgfsetbuttcap
\pgfsetmiterjoin
\pgfsetlinewidth{0.100000\du}
\pgfsetbuttcap
\pgfsetmiterjoin
\pgfsetdash{}{0pt}
\definecolor{dialinecolor}{rgb}{1.000000, 1.000000, 1.000000}
\pgfsetfillcolor{dialinecolor}
\pgfpathellipse{\pgfpoint{8.930800\du}{9.758660\du}}{\pgfpoint{0.127410\du}{0\du
}}{\pgfpoint{0\du}{0.127410\du}}
\pgfusepath{fill}
\definecolor{dialinecolor}{rgb}{0.000000, 0.000000, 0.000000}
\pgfsetstrokecolor{dialinecolor}
\pgfpathellipse{\pgfpoint{8.930800\du}{9.758660\du}}{\pgfpoint{0.127410\du}{0\du
}}{\pgfpoint{0\du}{0.127410\du}}
\pgfusepath{stroke}
\pgfsetbuttcap
\pgfsetmiterjoin
\pgfsetdash{}{0pt}
\definecolor{dialinecolor}{rgb}{0.000000, 0.000000, 0.000000}
\pgfsetstrokecolor{dialinecolor}
\pgfpathellipse{\pgfpoint{8.930800\du}{9.758660\du}}{\pgfpoint{0.127410\du}{0\du
}}{\pgfpoint{0\du}{0.127410\du}}
\pgfusepath{stroke}
\pgfsetlinewidth{0.100000\du}
\pgfsetdash{}{0pt}
\pgfsetdash{}{0pt}
\pgfsetbuttcap
\pgfsetmiterjoin
\pgfsetlinewidth{0.100000\du}
\pgfsetbuttcap
\pgfsetmiterjoin
\pgfsetdash{}{0pt}
\definecolor{dialinecolor}{rgb}{1.000000, 1.000000, 1.000000}
\pgfsetfillcolor{dialinecolor}
\pgfpathellipse{\pgfpoint{5.999550\du}{4.996160\du}}{\pgfpoint{0.127410\du}{0\du
}}{\pgfpoint{0\du}{0.127410\du}}
\pgfusepath{fill}
\definecolor{dialinecolor}{rgb}{0.000000, 0.000000, 0.000000}
\pgfsetstrokecolor{dialinecolor}
\pgfpathellipse{\pgfpoint{5.999550\du}{4.996160\du}}{\pgfpoint{0.127410\du}{0\du
}}{\pgfpoint{0\du}{0.127410\du}}
\pgfusepath{stroke}
\pgfsetbuttcap
\pgfsetmiterjoin
\pgfsetdash{}{0pt}
\definecolor{dialinecolor}{rgb}{0.000000, 0.000000, 0.000000}
\pgfsetstrokecolor{dialinecolor}
\pgfpathellipse{\pgfpoint{5.999550\du}{4.996160\du}}{\pgfpoint{0.127410\du}{0\du
}}{\pgfpoint{0\du}{0.127410\du}}
\pgfusepath{stroke}
\pgfsetlinewidth{0.100000\du}
\pgfsetdash{}{0pt}
\pgfsetdash{}{0pt}
\pgfsetbuttcap
\pgfsetmiterjoin
\pgfsetlinewidth{0.100000\du}
\pgfsetbuttcap
\pgfsetmiterjoin
\pgfsetdash{}{0pt}
\definecolor{dialinecolor}{rgb}{1.000000, 1.000000, 1.000000}
\pgfsetfillcolor{dialinecolor}
\pgfpathellipse{\pgfpoint{10.968300\du}{3.564910\du}}{\pgfpoint{0.127410\du}{
0\du}}{\pgfpoint{0\du}{0.127410\du}}
\pgfusepath{fill}
\definecolor{dialinecolor}{rgb}{0.000000, 0.000000, 0.000000}
\pgfsetstrokecolor{dialinecolor}
\pgfpathellipse{\pgfpoint{10.968300\du}{3.564910\du}}{\pgfpoint{0.127410\du}{
0\du}}{\pgfpoint{0\du}{0.127410\du}}
\pgfusepath{stroke}
\pgfsetbuttcap
\pgfsetmiterjoin
\pgfsetdash{}{0pt}
\definecolor{dialinecolor}{rgb}{0.000000, 0.000000, 0.000000}
\pgfsetstrokecolor{dialinecolor}
\pgfpathellipse{\pgfpoint{10.968300\du}{3.564910\du}}{\pgfpoint{0.127410\du}{
0\du}}{\pgfpoint{0\du}{0.127410\du}}
\pgfusepath{stroke}
\pgfsetlinewidth{0.100000\du}
\pgfsetdash{}{0pt}
\pgfsetdash{}{0pt}
\pgfsetmiterjoin
\pgfsetbuttcap
\definecolor{dialinecolor}{rgb}{1.000000, 1.000000, 1.000000}
\pgfsetfillcolor{dialinecolor}
\fill
(20.172886\du,4.978608\du)--(25.145026\du,3.559858\du)--(23.145026\du,
9.747358\du)--(18.172886\du,10.978608\du)--cycle;
\definecolor{dialinecolor}{rgb}{0.000000, 0.000000, 0.000000}
\pgfsetstrokecolor{dialinecolor}
\draw
(20.172886\du,4.978608\du)--(25.145026\du,3.559858\du)--(23.145026\du,
9.747358\du)--(18.172886\du,10.978608\du)--cycle;
\pgfsetlinewidth{0.100000\du}
\pgfsetdash{{\pgflinewidth}{0.200000\du}}{0cm}
\pgfsetdash{{\pgflinewidth}{0.200000\du}}{0cm}
\pgfsetbuttcap
{
\definecolor{dialinecolor}{rgb}{0.000000, 0.000000, 0.000000}
\pgfsetfillcolor{dialinecolor}
\definecolor{dialinecolor}{rgb}{0.000000, 0.000000, 0.000000}
\pgfsetstrokecolor{dialinecolor}
\draw (25.557626\du,9.159375\du)--(17.082626\du,11.259375\du);
}
\pgfsetlinewidth{0.100000\du}
\pgfsetdash{{\pgflinewidth}{0.200000\du}}{0cm}
\pgfsetdash{{\pgflinewidth}{0.200000\du}}{0cm}
\pgfsetbuttcap
{
\definecolor{dialinecolor}{rgb}{0.000000, 0.000000, 0.000000}
\pgfsetfillcolor{dialinecolor}
\definecolor{dialinecolor}{rgb}{0.000000, 0.000000, 0.000000}
\pgfsetstrokecolor{dialinecolor}
\draw (26.607626\du,3.150000\du)--(18.432626\du,5.437500\du);
}
\pgfsetlinewidth{0.050000\du}
\pgfsetdash{}{0pt}
\pgfsetdash{}{0pt}
\pgfsetbuttcap
\pgfsetmiterjoin
\pgfsetlinewidth{0.050000\du}
\pgfsetbuttcap
\pgfsetmiterjoin
\pgfsetdash{}{0pt}
\definecolor{dialinecolor}{rgb}{0.000000, 0.000000, 0.000000}
\pgfsetfillcolor{dialinecolor}
\pgfpathellipse{\pgfpoint{20.584633\du}{10.780856\du}}{\pgfpoint{0.111748\du}{
0\du}}{\pgfpoint{0\du}{0.111748\du}}
\pgfusepath{fill}
\definecolor{dialinecolor}{rgb}{0.000000, 0.000000, 0.000000}
\pgfsetstrokecolor{dialinecolor}
\pgfpathellipse{\pgfpoint{20.584633\du}{10.780856\du}}{\pgfpoint{0.111748\du}{
0\du}}{\pgfpoint{0\du}{0.111748\du}}
\pgfusepath{stroke}
\pgfsetbuttcap
\pgfsetmiterjoin
\pgfsetdash{}{0pt}
\definecolor{dialinecolor}{rgb}{0.000000, 0.000000, 0.000000}
\pgfsetstrokecolor{dialinecolor}
\pgfpathellipse{\pgfpoint{20.584633\du}{10.780856\du}}{\pgfpoint{0.111748\du}{
0\du}}{\pgfpoint{0\du}{0.111748\du}}
\pgfusepath{stroke}
\pgfsetlinewidth{0.100000\du}
\pgfsetdash{}{0pt}
\pgfsetdash{}{0pt}
\pgfsetbuttcap
\pgfsetmiterjoin
\pgfsetlinewidth{0.100000\du}
\pgfsetbuttcap
\pgfsetmiterjoin
\pgfsetdash{}{0pt}
\definecolor{dialinecolor}{rgb}{1.000000, 1.000000, 1.000000}
\pgfsetfillcolor{dialinecolor}
\pgfpathellipse{\pgfpoint{18.155296\du}{10.951018\du}}{\pgfpoint{0.127410\du}{
0\du}}{\pgfpoint{0\du}{0.127410\du}}
\pgfusepath{fill}
\definecolor{dialinecolor}{rgb}{0.000000, 0.000000, 0.000000}
\pgfsetstrokecolor{dialinecolor}
\pgfpathellipse{\pgfpoint{18.155296\du}{10.951018\du}}{\pgfpoint{0.127410\du}{
0\du}}{\pgfpoint{0\du}{0.127410\du}}
\pgfusepath{stroke}
\pgfsetbuttcap
\pgfsetmiterjoin
\pgfsetdash{}{0pt}
\definecolor{dialinecolor}{rgb}{0.000000, 0.000000, 0.000000}
\pgfsetstrokecolor{dialinecolor}
\pgfpathellipse{\pgfpoint{18.155296\du}{10.951018\du}}{\pgfpoint{0.127410\du}{
0\du}}{\pgfpoint{0\du}{0.127410\du}}
\pgfusepath{stroke}
\definecolor{dialinecolor}{rgb}{0.000000, 0.000000, 0.000000}
\pgfsetstrokecolor{dialinecolor}
\node[anchor=west] at (25.897886\du,2.991108\du){$\mathsmaller{F_1(1)}$};
\definecolor{dialinecolor}{rgb}{0.000000, 0.000000, 0.000000}
\pgfsetstrokecolor{dialinecolor}
\node[anchor=west] at (24.887886\du,9.136108\du){$\mathsmaller{F_1(-1)}$};
\pgfsetlinewidth{0.050000\du}
\pgfsetdash{}{0pt}
\pgfsetdash{}{0pt}
\pgfsetbuttcap
{
\definecolor{dialinecolor}{rgb}{0.000000, 0.000000, 0.000000}
\pgfsetfillcolor{dialinecolor}
\definecolor{dialinecolor}{rgb}{0.000000, 0.000000, 0.000000}
\pgfsetstrokecolor{dialinecolor}
\draw (20.650378\du,11.040625\du)--(18.787878\du,3.890625\du);
}
\definecolor{dialinecolor}{rgb}{0.498039, 0.498039, 0.498039}
\pgfsetfillcolor{dialinecolor}
\pgfpathellipse{\pgfpoint{19.107526\du}{5.225483\du}}{\pgfpoint{0.150000\du}{
0\du}}{\pgfpoint{0\du}{0.150000\du}}
\pgfusepath{fill}
\pgfsetlinewidth{0.100000\du}
\pgfsetdash{}{0pt}
\pgfsetdash{}{0pt}
\definecolor{dialinecolor}{rgb}{0.498039, 0.498039, 0.498039}
\pgfsetstrokecolor{dialinecolor}
\pgfpathellipse{\pgfpoint{19.107526\du}{5.225483\du}}{\pgfpoint{0.150000\du}{
0\du}}{\pgfpoint{0\du}{0.150000\du}}
\pgfusepath{stroke}
\definecolor{dialinecolor}{rgb}{0.498039, 0.498039, 0.498039}
\pgfsetfillcolor{dialinecolor}
\pgfpathellipse{\pgfpoint{20.500378\du}{10.403125\du}}{\pgfpoint{0.150000\du}{
0\du}}{\pgfpoint{0\du}{0.150000\du}}
\pgfusepath{fill}
\pgfsetlinewidth{0.100000\du}
\pgfsetdash{}{0pt}
\pgfsetdash{}{0pt}
\definecolor{dialinecolor}{rgb}{0.498039, 0.498039, 0.498039}
\pgfsetstrokecolor{dialinecolor}
\pgfpathellipse{\pgfpoint{20.500378\du}{10.403125\du}}{\pgfpoint{0.150000\du}{
0\du}}{\pgfpoint{0\du}{0.150000\du}}
\pgfusepath{stroke}
\definecolor{dialinecolor}{rgb}{0.000000, 0.000000, 0.000000}
\pgfsetstrokecolor{dialinecolor}
\node[anchor=west] at
(20.500378\du,10.703125\du){$\mathsmaller{\mbox{\smaller{proj}}_{F_1(-1)}(y)}$};
\definecolor{dialinecolor}{rgb}{0.000000, 0.000000, 0.000000}
\pgfsetstrokecolor{dialinecolor}
\node[anchor=west] at
(19.107526\du,5.625483\du){$\mathsmaller{\mbox{\smaller{proj}}_{F_1(1)}(y)}$};
\definecolor{dialinecolor}{rgb}{0.000000, 0.000000, 0.000000}
\pgfsetstrokecolor{dialinecolor}
\node[anchor=west] at (19.979136\du,11.422408\du){$\mathsmaller{y}$};
\pgfsetlinewidth{0.100000\du}
\pgfsetdash{}{0pt}
\pgfsetdash{}{0pt}
\pgfsetbuttcap
\pgfsetmiterjoin
\pgfsetlinewidth{0.100000\du}
\pgfsetbuttcap
\pgfsetmiterjoin
\pgfsetdash{}{0pt}
\definecolor{dialinecolor}{rgb}{1.000000, 1.000000, 1.000000}
\pgfsetfillcolor{dialinecolor}
\pgfpathellipse{\pgfpoint{23.103686\du}{9.737268\du}}{\pgfpoint{0.127410\du}{
0\du}}{\pgfpoint{0\du}{0.127410\du}}
\pgfusepath{fill}
\definecolor{dialinecolor}{rgb}{0.000000, 0.000000, 0.000000}
\pgfsetstrokecolor{dialinecolor}
\pgfpathellipse{\pgfpoint{23.103686\du}{9.737268\du}}{\pgfpoint{0.127410\du}{
0\du}}{\pgfpoint{0\du}{0.127410\du}}
\pgfusepath{stroke}
\pgfsetbuttcap
\pgfsetmiterjoin
\pgfsetdash{}{0pt}
\definecolor{dialinecolor}{rgb}{0.000000, 0.000000, 0.000000}
\pgfsetstrokecolor{dialinecolor}
\pgfpathellipse{\pgfpoint{23.103686\du}{9.737268\du}}{\pgfpoint{0.127410\du}{
0\du}}{\pgfpoint{0\du}{0.127410\du}}
\pgfusepath{stroke}
\pgfsetlinewidth{0.100000\du}
\pgfsetdash{}{0pt}
\pgfsetdash{}{0pt}
\pgfsetbuttcap
\pgfsetmiterjoin
\pgfsetlinewidth{0.100000\du}
\pgfsetbuttcap
\pgfsetmiterjoin
\pgfsetdash{}{0pt}
\definecolor{dialinecolor}{rgb}{1.000000, 1.000000, 1.000000}
\pgfsetfillcolor{dialinecolor}
\pgfpathellipse{\pgfpoint{20.172436\du}{4.974768\du}}{\pgfpoint{0.127410\du}{
0\du}}{\pgfpoint{0\du}{0.127410\du}}
\pgfusepath{fill}
\definecolor{dialinecolor}{rgb}{0.000000, 0.000000, 0.000000}
\pgfsetstrokecolor{dialinecolor}
\pgfpathellipse{\pgfpoint{20.172436\du}{4.974768\du}}{\pgfpoint{0.127410\du}{
0\du}}{\pgfpoint{0\du}{0.127410\du}}
\pgfusepath{stroke}
\pgfsetbuttcap
\pgfsetmiterjoin
\pgfsetdash{}{0pt}
\definecolor{dialinecolor}{rgb}{0.000000, 0.000000, 0.000000}
\pgfsetstrokecolor{dialinecolor}
\pgfpathellipse{\pgfpoint{20.172436\du}{4.974768\du}}{\pgfpoint{0.127410\du}{
0\du}}{\pgfpoint{0\du}{0.127410\du}}
\pgfusepath{stroke}
\pgfsetlinewidth{0.100000\du}
\pgfsetdash{}{0pt}
\pgfsetdash{}{0pt}
\pgfsetbuttcap
\pgfsetmiterjoin
\pgfsetlinewidth{0.100000\du}
\pgfsetbuttcap
\pgfsetmiterjoin
\pgfsetdash{}{0pt}
\definecolor{dialinecolor}{rgb}{1.000000, 1.000000, 1.000000}
\pgfsetfillcolor{dialinecolor}
\pgfpathellipse{\pgfpoint{25.141186\du}{3.543518\du}}{\pgfpoint{0.127410\du}{
0\du}}{\pgfpoint{0\du}{0.127410\du}}
\pgfusepath{fill}
\definecolor{dialinecolor}{rgb}{0.000000, 0.000000, 0.000000}
\pgfsetstrokecolor{dialinecolor}
\pgfpathellipse{\pgfpoint{25.141186\du}{3.543518\du}}{\pgfpoint{0.127410\du}{
0\du}}{\pgfpoint{0\du}{0.127410\du}}
\pgfusepath{stroke}
\pgfsetbuttcap
\pgfsetmiterjoin
\pgfsetdash{}{0pt}
\definecolor{dialinecolor}{rgb}{0.000000, 0.000000, 0.000000}
\pgfsetstrokecolor{dialinecolor}
\pgfpathellipse{\pgfpoint{25.141186\du}{3.543518\du}}{\pgfpoint{0.127410\du}{
0\du}}{\pgfpoint{0\du}{0.127410\du}}
\pgfusepath{stroke}
\definecolor{dialinecolor}{rgb}{0.000000, 0.000000, 0.000000}
\pgfsetstrokecolor{dialinecolor}
\node[anchor=west] at (22.007626\du,3.225000\du){};
\pgfsetlinewidth{0.070000\du}
\pgfsetdash{{\pgflinewidth}{0.200000\du}}{0cm}
\pgfsetdash{{\pgflinewidth}{0.200000\du}}{0cm}
\definecolor{dialinecolor}{rgb}{0.000000, 0.000000, 0.000000}
\pgfsetstrokecolor{dialinecolor}
\pgfpathellipse{\pgfpoint{6.384640\du}{10.778130\du}}{\pgfpoint{2.437500\du}{
0\du}}{\pgfpoint{0\du}{2.437500\du}}
\pgfusepath{stroke}
\definecolor{dialinecolor}{rgb}{0.000000, 0.000000, 0.000000}
\pgfsetstrokecolor{dialinecolor}
\node[anchor=west] at (2.084640\du,18.907800\du){\smaller{(a) H - original
column ordering}};
\definecolor{dialinecolor}{rgb}{0.000000, 0.000000, 0.000000}
\pgfsetstrokecolor{dialinecolor}
\node[anchor=west] at (16.234600\du,18.907800\du){\smaller{(b) H - columns
swapped}};
\pgfsetlinewidth{0.070000\du}
\pgfsetdash{{\pgflinewidth}{0.200000\du}}{0cm}
\pgfsetdash{{\pgflinewidth}{0.200000\du}}{0cm}
\definecolor{dialinecolor}{rgb}{0.000000, 0.000000, 0.000000}
\pgfsetstrokecolor{dialinecolor}
\pgfpathellipse{\pgfpoint{20.604128\du}{10.800625\du}}{\pgfpoint{2.437500\du}{
0\du}}{\pgfpoint{0\du}{2.437500\du}}
\pgfusepath{stroke}
\end{tikzpicture}

\caption{Geometry of the search  with two different column ordering.}
\label{SEGeometry}
\end{figure}
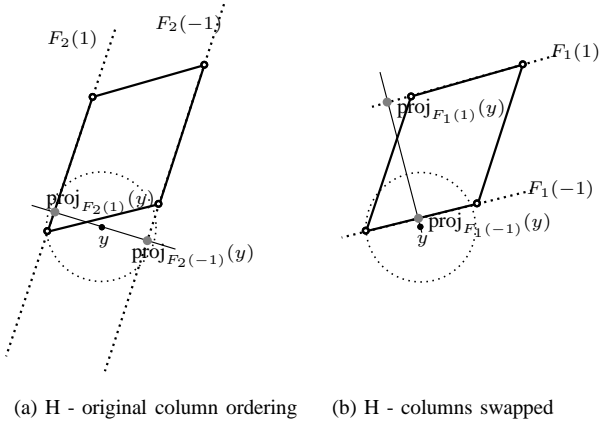

Fig.\ \ref{SEGeometry} shows a 2-D BILS problem;
\ref{SEGeometry}(a) represents the original column ordering and 
\ref{SEGeometry}(b) is after the columns have been swapped.

In the SW algorithm $H=[h_1,\ldots, h_n]$ is assumed to be square and non-singular.
Let 
$$
G =[g_1,\ldots, g_n]= H^{-T}.
$$ 
For any  integer $\alpha$,   \cite{SuW05} defines the
affine sets, $F_i(\alpha) = \{w \ | \ g_i^T(w-h_i\alpha) = 0\}$.
The lattice points generated by $H$ occur at the intersections of these affine sets. 
Let the orthogonal projection of a vector $s$ onto a vector $t$ be denoted as
$\mbox{proj}_t(s)$, then 
the orthogonal projection of some vector $s$ onto $F_i(\alpha)$ is 
$\mbox{proj}_{F_i(\alpha)}(s) = s -\mbox{proj}_{g_i}(s-h_i\alpha).$ 
Therefore the orthogonal distance between $s$ and
$F_i(\alpha)$ is $\dist(s,F_i(\alpha)) =  \| s - \mbox{proj}_{F_i(\alpha)}(s) \|_2$. 
In \cite{SuW05}, the points labeled $\mbox{proj}_{F_2(1)}(y)$ and
$\mbox{proj}_{F_2(-1)}(y)$ in Fig.\  \ref{SEGeometry}
are called residual targets  and ``represent the
components [of $y$] that remain after an orthogonal part has been projected
away.''  

Note that $F_2(\alpha)$ in Fig.\ \ref{SEGeometry} is a sublattice of dimension $1$.  
Algebraically it is the lattice generated by $H$ with column
$2$ removed. It can also be thought of as a subtree of the search tree where
$x_2 = \alpha$ has been fixed. 
In the first step of the search process for a general case,   $x_n$ is chosen to be
 $x_n=\arg\min_{\alpha\in {\cal B}_n}\dist(y,F_n(\alpha))$;
 thus $F_n(x_n)$ is the nearest affine set to $y$. 
Actually the value of $x_n$ is identical to $\lfloor c_n \rceil_{{\cal B}_n}$ given in Section ~\ref{sec:Search},
which will be proved later.
Then   $y$ is updated  as $y := \mbox{proj}_{F_n(x_n)}(y) - h_nx_n$. If we look
at Fig.\ \ref{SEGeometry}, we see that the projection $\mbox{proj}_{F_n(x_n)}(y)$ moves $y$ onto $F_n(x_n)$, while the subtraction of $h_nx_n$ algebraically fixes the value of $x_n$. This is necessary because in subsequent steps we will not consider the column $h_n$.

We now apply the same process to the new $n-1$ dimensional search space
$F_n(x_n)$. If at some level $i$, $\min_{\alpha\in {\cal B}_i}\dist(y,F_i(\alpha))$ exceeds the current
search radius, we must move back to level $i+1$. 
When the search process reaches level $1$ and fixes $x_1$, it updates the radius to  
$\dist(y,F_1(x_1))$ and moves back up to level $2$.

Note that this search process is mathematically equivalent to the one described in section
\ref{sec:Search}; the difference is that it  does projections
because  the generator matrix is not assumed to be upper-triangular. 
Computationally the former is more expensive than the latter.

To see the motivation of the SW algorithm for choosing a particular column ordering,
consider Fig.\ \ref{SEGeometry}. Suppose the search algorithm has knowledge of
the residual for the optimal solution (the radius of the circle in the diagram).
With the column ordering chosen in (a), there are two possible choices for $x_2$,
leading to the two dashed lines $F_2(-1)$ and $F_2(1)$ which cross the circle. This means
that we will need to find $x_1$ for both of these choices
before we can determine which one leads to the optimum solution. In (b), there
is only one possible choice for $x_1$,  leading to the only dashed line $F_1(-1)$
which crosses the circle, meaning we only need to find $x_2$ to find the optimum solution.
Since the projection resulting from the correct choice of $x_2$ will always be
within the sphere, it makes sense to choose the ordering which maximizes the
distance to the second best choice for $x_2$ in hopes that the second nearest
choice will result in a value for $\min_{\alpha\in {\cal B}_2}\dist(y,F_2(\alpha))$ outside the sphere and the
dimensionality can be reduced by one. 
For more detail on the geometry, see  \cite{SuW05}.

The following will give an overview of the SW algorithm as given in
\cite{SuW05} but described in a framework similar to what was used to describe
CH. In the first step to determine the last column, for each $i = 1, \dots, n$,
we compute 
\be
x_i^s \!=\! \arg\min_{\alpha\in {\cal B}_i}\dist(y,F_i(\alpha)) 
\!=\!  \arg\min _{\alpha\in {\cal B}_i}| y^Tg_i - \alpha | \!=\! \left \lfloor y^Tg_i \right \rceil_{{\cal B}_i},
\label{eq:xis}
\ee
where the superscript $s$  stands for the SW algorithm.
Let $\bar{x}_i^s$ be the second closest integer in ${{\cal B}_i }$ to $y^Tg_i$, i.e.,
$\bar{x}_i^s = \left \lfloor y^Tg_i \right \rceil_{{\cal B}_i\backslash x_i^s}.$
Let $j = \arg\max_i  \dist (y,F_i(\bar{x}_i^s))$. 
Then SW chooses column $j$ as the last column of the final reordered $H$,
 updates $y$ by setting 
$y:=\mbox{proj}_{F_j(x_j^s)}(y) -h_jx_j^s$  and 
updates $G$ by setting $ g_i: = \mbox{proj}_{F_j(0)}(g_i)$ for all $i\neq j$. 
After $G$ and $y$ have been updated, the algorithm continues to find column $n-1$ in the
same way etc. The pseudo-code of the SW algorithm is given in Algorithm \ref{alg:SWOrig}.
\begin{algorithm}
\caption{SW Algorithm - Returns $p$, the column permutation vector}
\label{alg:SWOrig}
\begin{algorithmic}[1]
\STATE $p := 1:n$
\STATE $p' := \{1, 2, \ldots, n\}$
\STATE \label{l:swG} $G := H^{-T}$ \hfill 
\FOR{$k=n$ to $2$}
	\STATE $maxDist := -1$
	\FOR{$i \in p'$}
		\STATE $x_i^s := \left \lfloor  y^Tg_i \right \rceil_{{\cal B }_i}$ \hfill
		     \label{l:swx}
		\STATE $\bar{x}_i^s := \left \lfloor y^Tg_i \right \rceil_{{{\cal B }_i}{\backslash x_i^s}}$
		    \label{l:swbx}
		\STATE  \label{l:swd} $\dist_i^s := \dist(y,F_i(\bar{x}_i^s))$ 
		\IF{$dist_i^s > maxDist$}
			\STATE $maxDist := dist_i^s$
			\STATE $j := i$
		\ENDIF
	\ENDFOR
	\STATE $p_k := j$
	\STATE $p' := p' \backslash j$
	\STATE $y := \mbox{proj}_{F_j(x_j^s)}(y) -h_jx_j^s$  \label{l:swy}
	\FOR{$i \in p'$}
		\STATE \label{l:swg} $g_i := \mbox{proj}_{F_j(0)}(g_i) $ 
	\ENDFOR
\ENDFOR
\STATE $p_1 := p'$
\end{algorithmic}
\end{algorithm}

 \cite{SuW05} did not say how to implement the algorithm and did not give a complexity analysis.
The parts of the cost we must consider for implementation occur in
lines \ref{l:swd} and \ref{l:swg}.
Note that $ \dist(y,F_i(\bar{x}_i^s))=\|\mbox{proj}_{g_i}(y-h_i\bar{x}_i^s)\|_2$
and $\mbox{proj}_{F_j(0)}(g_i)= g_i -\mbox{proj}_{g_i} g_i$,
where $\mbox{proj}_{g_i}=g_ig_i^\dag = g_ig_i^T/\|g_i\|^2$. 
A naive implementation would first compute $\mbox{proj}_{g_i}$, requiring $n^2$ flops, then compute 
$\|\mbox{proj}_{g_i}(y-h_i\bar{x}_i^s)\|_2$ and $g_i -\mbox{proj}_{g_i} g_i$, each requiring $2n^2$ flops.
Summing these costs over all loop iterations we get a total complexity of $2.5n^4$ flops.
In the next subsection we will simplify some steps in Algorithm \ref{alg:SWOrig}
and show how to  implement them efficiently.

\subsection{Algebraic Interpretation and Modifications of SW}
\label{sec:improvedSW}
In this section we give new algebraic interpretation of some steps in Algorithm 2,
simplify some key steps to improve the efficiency,
and  extend the algorithm to handle a more general case.
All line numbers refer to Algorithm 2.

First we show  how to efficiently compute $\dist_i^s$ in line \ref{l:swd}. 
Observing that $g_i^Th_i = 1$, we have
\begin{equation}
\label{eq:newDist}
\dist_i^s =   \|  g_ig_i^\dag (y-h_i\bar{x}_i^s)  \|_2 
=    | y^Tg_i -\bar{x}_i^s |/\| g_i  \|_2.
\end{equation} 
Note that $y^Tg_i$ and $\bar{x}_i^s$ have been computed in lines \ref{l:swx} and \ref{l:swbx}, respectively.
So the main cost of computing $\dist_i^s$ is the cost of computing $\|g_i\|_2$,
requiring only $2n$ flops. 
For $k=n$ in Algorithm 2,  $y^Tg_i=y^TH^{-T}e_i=(H^{-1}y)^Te_i$, i.e.,  $y^Tg_i$
is the $i^{th}$ entry of the real solution for $Hx=y$. 
The interpretation can be generalized to  a general $k$. 

In line \ref{l:swg} Algorithm 2,  
\begin{align}
g_i^{\small \mbox{new}} & \equiv \mbox{proj}_{F_j(0)}(g_i)   \nonumber \\
  & =(I- \mbox{proj}_{g_j})g_i=g_i- g_j(g_j^Tg_i/\|g_j\|_2^2). \label{eq:gup}
\end{align}
Using the last expression for computation needs only $4n$ flops
(note that $\|g_j\|_2$ has been computed before, see \eqref{eq:newDist}).
We can actually show that the above is performing  updating of $G$, the Moore-Penrose generalized inverse of
$H$ after we remove its $j^{th}$ column. 
For proof of this, see \cite{Cli64}.

In line \ref{l:swy} of Algorithm 2,
\begin{align}
y^{\small \mbox{new}} & \!\equiv\! \mbox{proj}_{F_j(x_j^s)}(y) - h_jx_j^s 
 \!=\! (y -  g_jg_j^\dag(y-h_jx_j^s)) - h_jx_j^s  \nonumber \\
   &=  (I-\mbox{proj}_{g_j})(y-h_jx_j^s). \label{eq:yup}  
\end{align}
This means that after $x_j$ is fixed to be $x_j^s$, $h_jx_j^s$ is combined with $y$ (the same
as CH does)  and then the vector is projected to the orthogonal complement of 
the space spanned by $g_j$. 
We can show that this guarantees that the updated $y$ is in the subspace spanned by
the columns of $H$ which have not been chosen.
This is consistent with the assumption that  $H$ is nonsingular, which implies that 
the original $y$  is in the space spanned by  the columns of $H$.
However, it is not necessary to apply the orthogonal projector $I- \mbox{proj}_{g_j}$ to $y-h_jx_j^s$ in \eqref{eq:yup}.
The reason is as follows. 
In Algorithm 2, $y^{\small \mbox{new}}$ and $g_i^{\small \mbox{new}}$ will be used only for computing 
$(y^{\small \mbox{new}})^Tg_i^{\small \mbox{new}}$ (see line \ref{l:swx}).
But from \eqref{eq:gup} and \eqref{eq:yup}
\begin{align*}
(y^{\small \mbox{new}})^Tg_i^{\small \mbox{new}}
& =(y-h_jx_j^s)^T(I-\mbox{proj}_{g_j})(I-\mbox{proj}_{g_j})g_i \\
&=(y-h_jx_j^s)^Tg_i^{\small \mbox{new}}.
\end{align*}
Therefore, line \ref{l:swy} can be replaced by $y:=y-h_jx_j^s$.
This not only simplifies the computation but also is much easier to interpret---after $x_j$ is fixed to be $x_j^s$,  
 $h_jx_j^s$ is combined into $y$ as what the CH algorithm does.
Let $H_{:,1:n-1}$ denote $H$ after its $j^{th}$ column is removed. 
We then continue to work on the subproblem
\be
\min_{\check{x}\in \mathbb{Z}^{n-1}}\|y-H_{:,1:n-1}\check{x}\|_2, 
\label{eq:subs}
\ee
where $\check{x}=[x_1,\ldots,x_{j-1},x_{j+1},\ldots,x_n]^T$ 
satisfies the corresponding box constraint.
Here $H_{:,1:n-1}$ is not square. But there is no problem to handle
it, see the next paragraph.

In \cite{SuW05},  $H$ is assumed to be square and non-singular. 
In our opinion, this condition may cause confusion,
since for each $k$ except $k=n$ in Algorithm 2, 
the remaining columns of $H$ which have not been chosen do not form a square matrix.
Also the condition restricts the application of the algorithm to a general full column rank matrix $H$,
unless we transform $H$ to a nonsingular matrix $R$ by the QR factorization.
To extend the algorithm to a general full column rank matrix $H$, we need only 
replace line \ref{l:swG} by $G:=(H^{\dagger})^T$.
This extension has another benefit. 
We mentioned before that the updating of $G$ in line \ref{l:swg}
is actually the updating of the Moore-Pernrose generalized inverse 
of the matrix formed by the columns of $H$ which have not been chosen. 
So the extension makes all steps consistent.

To reliably compute $G$ for a general full column rank $H$,
we can compute the  QR factorization $H=Q_1R$ by the Householder transformations
and then solve the triangular system  $RG^T=Q_1^T$ to obtain $G$.
This requires $(5m-4n/3)n^2$ flops. 
Another less reliable but more efficient way to do this is to compute $G=H(H^TH)^{-1}$. 
To do this efficiently we would compute the Cholesky factorization  $H^TH = R^TR$ and solve 
$R^TRG^T = H^T$ for $G$ by using the triangular structure of $R$. 
The total cost for computing $G$ by this method can be shown to be $3mn^2+\frac{n^3}{3}$.
If $H$ is square and nonsingular, we would use the LU factorization with partial pivoting to compute $H^{-1}$
and the cost is $2n^3$ flops.

For the rest part of the algorithm if we use the simplification and efficient implementations
mentioned above, we can show that it needs $4mn^2$ flops. 

We see the modified SW algorithm is much more efficient than both the CH algorithm
and the SW algorithm implemented in a naive way we mentioned in the previous subsection.

\section{Equivalence of CH and SW}\label{sec:equivalence}
In this section we prove that  CH and  the modified  SW produce the same set of permutations
for a general full column rank $H$.
To prove this it will suffice to prove that $x_i^s = x_i^c$, $\bar{x}_i^s =\bar{x}_i^c$,
$\dist_i^s = \dist_i^c$ for $i=1, \ldots, n$ in the first step which determines the last column of the final reordered $H$
and that the subproblems produced for the second step of
each algorithm are equivalent. 

Proving $x_i^s = x_i^c$ is not difficult.
The only effect the interchange of columns $i$  and $n$ of $R$ in CH  
has on the real LS solution is that elements $i$ and $n$ of the solution are swapped.
Therefore $x_i^c$ is just the $i^{th}$ element of the real LS
solution rounded to the nearest integer in ${{\cal B}_i }$. 
Thus, with \eqref{eq:xic} and \eqref{eq:xis},
\be
x_i^c=   \lfloor (H^{\dagger}y)_i  \rceil_{{\cal B}_i }
=  \lfloor e_i^T H^{\dagger}y   \rceil_{{\cal B}_i }
=  \lfloor g_i^T  y \rceil_{{\cal B}_i } =x_i^s.
\label{eq:xics}
\ee
Therefore we also have $\bar{x}_i^c=\bar{x}_i^s$.


In CH, after applying a permutation $P$ to swap columns $i$ and $n$ of $R$,  
we apply $V^T$, a product of the Givens rotations, to bring $R$ back to a new upper triangular
matrix, denoted by $\hat{R}$, and also apply $V$ to $\bar{y}$, 
leading to  $\hat{y} = V^T\bar{y}$. 
Thus  $\hat{R}=V^T RP$ and $\hat{y} = V^T\bar{y}=V^TQ_1^Ty$. 
Then $H=Q_1R= Q_1V\hat{R}P^T$, $H^\dag= P\hat{R}^{-1}V^TQ_1^T$, 
$g_i=(H^\dag)^Te_i=Q_1V\hat{R}^{-T}P^Te_i=Q_1V\hat{R}^{-T}e_n$,
and $\|g_i\|_2=\|\hat{R}^{-T}e_n\|_2=1/|\hat{r}_{nn}|$.
Therefore, with \eqref{eq:newDist} and \eqref{eq:dic}
\begin{align}
\dist_i^s
&=\frac{ | y^Tg_i - \bar{x}_i^s   |}{  \| g_i   \|_2} 
=|\hat{r}_{nn}||y^TQ_1V\hat{R}^{-T}e_n- \bar{x}_i^s  |  \label{eq:disc} \\
& = |\hat{r}_{nn}|| \hat{y}_n/\hat{r}_{nn} - \bar{x}_i^s | 
 = |\hat{r}_{nn}(c_n-\bar{x}_i^c)| =\dist_i^c.  \nonumber
\end{align}

Now we consider  the subproblem \eqref{eq:subc} in CH and the subproblem \eqref{eq:subs} in SW.
We can easily show that $R_{1:n-1,1:n-1}$ in  \eqref{eq:subc} is the $R$-factor of the QR factorization
of $H_{:,1:n-1}P$, where $H_{:,1:n-1}$ is the matrix given in \eqref{eq:subs}
and $P$ is a permutation matrix such that $\check{x}=P\tilde{x}$,
and that $\bar{y}_{1:n-1}$ in  \eqref{eq:subc} is the multiplication of the transpose of 
the $Q_1$-factor of the QR factorization of $H_{:,1:n-1}P$ and $y$ in \eqref{eq:subs}.
Thus the two subproblems are equivalent.

\section{New Algorithm}

Now that we know the two algorithms are equivalent, we can take the best
parts from both and combine them to form a new algorithm. 
The main cost in CH is to interchange the columns of $R$ and return it to
upper-triangular form using Givens rotations. 
When we determine the $k^{th}$ column,  we must do this $k$ times. 
We can avoid all but one of these column interchanges by computing $x_i^c$, 
$\bar{x}_i^c$ and $\dist_i^c$ directly. 

After the QR factorization of $H$, we  solve the reduced ILS problem \eqref{eq:bils}.
We need only consider how to determine the last column of the final $R$.
Other columns can be determined similarly. 
Here we use the ideas from SW.
Let $G=R^{-T}$, which is lower triangular.
By \eqref{eq:xics}, we compute for $i=1,\ldots, n$
\begin{align*}
& x_i = \left \lfloor \bar{y}^TG_{:,i} \right \rceil_{{\cal B}_i} 
=\left \lfloor \bar{y}_{i:n}^T G_{i:n,i} \right \rceil_{{\cal B}_i}, \;
\bar{x}_i=\left \lfloor \bar{y}_{i:n}^T G_{i:n,i} \right \rceil_{{\cal B}_i\backslash x_i}, \\
& \dist_i = |\bar{y}_{i:n}^T G_{i:n,i}-\bar{x}_i|/\|G_{i:n,i}\|_2.
\end{align*}

Let $j=\arg\max_{i} \dist_i$. We  take a slightly different approach to
permuting the columns than was used in CH. Once $j$ is determined, we  
set $\bar{y}_{1:n-1} := \bar{y}_{1:n-1} - r_{1:n-1,j}x_j$. Then we simply remove
the $j^{th}$ column from $R$, and restore it to upper triangular using
Givens rotations. We then apply the same Givens rotations to the new
$\bar{y}$. In addition, we must also update the inverse matrix $G$. This is
very easy, we can just remove the $j^{th}$ column of $G$ and apply the same
Givens rotations that were used to restore the upper triangular structure of
$R$. To see this is true notice that removing column $j$ of $R$ is
mathematically equivalent to rotating $j$ to the last column and shifting
columns $j, j+1, \ldots, n$ to the left one position, since we will only consider columns
$1, 2, \ldots, n-1$ in subsequent steps. Suppose $P$ is the permutation matrix which
will permute the columns as described, and $V^T$ is the product of Givens
rotations to restore $R$ to upper-triangular. Let $\hat{R} = V^TRP$ and set
$\hat{G} = \hat{R}^{-T}$. Then
$$
\hat{G} = (V^TRP)^{-T} = V^TR^{-T}P = V^TGP.
$$
This indicates that  the same $V$ and $P$, which are used to transform $R$ to  $\hat{R}$, 
also transform $G$ to $\hat{G}$.
Since $\hat{G}$ is lower triangular, it is easy to verify that
$\hat{G}_{1:n-1,1:n-1} = \hat{R}^{-T}_{1:n-1,1:n-1}$.
Both $\hat{R}_{1:n-1,1:n-1}$ and $\hat{G}_{1:n-1,1:n-1}$ will be used in the next step.

After this, as in the CH algorithm, we continue to work on the subproblem of size $n-1$. 
The advantages of using the ideas from CH are that we always have a lower triangular $G$
whose dimension is reduced by one at each step
and the updating of $G$ is numerically stable as we use orthogonal transformations.
We give the pseudocode of the new algorithm in Algorithm  \ref{alg:NEW}.

\begin{algorithm}
\caption{New algorithm}
\label{alg:NEW}
\begin{algorithmic}[1]
\STATE  Compute the QR factorization of $H$ by Householder transformations: 
$\bsmx Q_1^T \\ Q_2^T \esmx H= \bsmx R\\ 0 \esmx$  \\
             and compute  $\bar{y} : = Q_1^Ty$ \hfill ($2(m-n/3)n^2$ flops)
\STATE $G := R^{-T}$ \hfill ($\frac{n^3}{3}$ flops)
\STATE $p := 1:n$
\STATE $p' := 1:n$
\FOR{$k=n$ to $2$}
	\STATE $maxDist := -1$
	\FOR{$i=1$ to $k$}
	         \STATE $\alpha=y_{i:k}^TG_{i:k,i}$
	         \STATE $x_i := \left \lfloor \alpha \right \rceil_{{\cal B}_i}$ \hfill ($2(k-i)$ flops)
	         \STATE $\bar{x}_i := \left \lfloor \alpha \right \rceil_{{{\cal B}_i}\backslash x_i}$
	         \STATE $\dist_i =|\alpha-\bar{x}_i|/ \| G_{i:k,i} \|_2$ \hfill ($2(k-i)$ flops)
			 \IF{$dist_i > maxDist$}
			 	\STATE $maxDist : = dist_i$
			 	\STATE $j:=i$
			 \ENDIF	
	\ENDFOR
	\STATE $p_k := p'_j$
	\STATE Interchange the intervals ${{\cal B}_k}$ and ${{\cal B}_j}$
	\STATE Interchange entries $k$ and $j$ in $p'$
	\STATE Set $\bar{y}:=\bar{y}_{1:k-1} - R_{1:k-1,j}x_j$	
	\STATE Remove column $j$ of $R$ and $G$, and return $R$ and $G$ to upper and lower triangular 
	by Givens rotations, respectively, and then remove the last row of $R$ and $G$. 
	The same Givens rotations are applied to $\bar{y}$. \\  \hfill ($6k(k-j)$ flops)
\ENDFOR
\STATE $p_1 = p'_1$
\end{algorithmic}
\end{algorithm}

Here we consider the complexity analysis of our new algorithm. 
If we sum the costs in algorithm \ref{alg:NEW} over all loop iterations,
we get a total of $\frac{7n^3}{3} + 2mn^2$ flops in the worst case. 
The worst case is very unlikely to occur, it arises when $j=1$ each iteration of the outer loop. In the average case
however, $j$ is around $k/2$ and we get an average case complexity of $\frac{4n^3}{3} + 2mn^2$ flops.
In both cases, the complexity is less than the complexity of the modified SW algorithm.

\section{Summary}
We showed that two algorithms for the column reordering  of the
box-constrained ILS problem are equivalent. To do that, we modified 
one algorithm and gave new insight.  
We proposed a new algorithm by combining
 the best ideas from both of the originals. Our new algorithm is more
efficient than either of the originals and is easy to implement and understand. 
Since the
three algorithms are theoretically equivalent and were derived through different
motivations, we now have both geometrical and algebraic motivations for the
algorithms.  

\bibliographystyle{IEEEtran}
\bibliography{IEEEabrv,./ILS}

%


\end{document}